\documentclass[conference]{IEEEtran}

\usepackage{graphicx}
\usepackage[font=small,labelfont=bf]{caption}
\usepackage[utf8]{inputenc} 
\usepackage[T1]{fontenc}    
\usepackage{hyperref}       
\usepackage{url}            
\usepackage{booktabs}       
\usepackage{amsfonts}       
\usepackage{nicefrac}       
\usepackage{microtype}      
\usepackage{lipsum}
\usepackage{authblk}
\usepackage[table]{xcolor}
\usepackage{svg}
\usepackage{amsmath}
\usepackage{tikz}
\usepackage{textcomp}
\usepackage{hyperref}

\DeclareCaptionLabelFormat{andtable}{#1~#2  \&  \tablename~\thetable}

\newcommand\copyrighttext{%
  \footnotesize \textcopyright 2020 IEEE.  Personal use of this material is permitted.  Permission from
IEEE must be obtained for all other uses, in any current or future media,
including reprinting/republishing this material for advertising or
promotional purposes, creating new collective works, for resale or
redistribution to servers or lists, or reuse of any copyrighted component
of this work in other works.}
\newcommand\copyrightnotice{%
\begin{tikzpicture}[remember picture,overlay]
\node[anchor=south,yshift=10pt] at (current page.south) {\fbox{\parbox{\dimexpr\textwidth-\fboxsep-\fboxrule\relax}{\copyrighttext}}};
\end{tikzpicture}%
}

\title{Latch-Based Logic Locking}

\author[1]{Joseph Sweeney}
\author[1]{Mohammed Zackriya V}
\author[2]{Samuel Pagliarini}
\author[1]{Lawrence Pileggi}
\affil[1]{Department of Electrical and Computer Engineering, Carnegie Mellon University}
\affil[2]{Centre for Hardware Security, Tallinn University of Technology}
\affil[ ]{\textit {Emails: \{joesweeney,mdzackriya,pileggi\}@cmu.edu}, samuel.pagliarini@taltech.ee}

\date{}                     
\begin{document}

\graphicspath{ {./images/} }
\maketitle
\copyrightnotice
\begin{abstract}
Globalization of IC manufacturing has led to increased security concerns, notably IP theft. Several logic locking techniques have been developed for protecting designs, but they typically display very large overhead, and are generally susceptible to deciphering attacks. In this paper, we propose latch-based logic locking, which manipulates both the flow of data and logic in the design. This method converts an interconnected subset of existing flip-flops to pairs of latches with programmable phase. In tandem, decoy latches and logic are added, inhibiting an attacker from determining the actual design functionality. To validate this technique, we developed and verified a locking insertion flow, analyzed PPA and ATPG overhead on benchmark circuits and industry cores, extended existing attacks to account for the technique, and taped out a demonstration chip. Importantly, we show that the design overhead with this approach is significantly less than with previous logic locking schemes, while resisting model checker-based, oracle-driven attacks. With minimal delay overhead, large numbers of decoy latches can be added, cheaply increasing attack resistance.
\end{abstract}


\section{Introduction}
Due to prohibitively high research and development costs, only a few foundries are manufacturing integrated circuits (ICs) in advanced technology nodes. Consequently, many IC companies tend to operate fabless, relying on untrusted foundries to manufacture their designs. Once a circuit is sent for fabrication, the foundry gains full visibility of the intellectual property (IP) in netlist form with minimal effort, allowing IP theft. This threat undermines the significant cost associated with developing digital circuits and is a growing concern in the IC industry \cite{Guin2017AOverproduction}.

To combat IP theft, a variety of logic locking techniques have been developed. These techniques add programmable elements to the logic of a circuit. When programmed incorrectly, the elements disrupt the functionality of the circuit, obfuscating the true design. The key, which correctly programs the elements, is stored in an on-chip, tamper-proof memory. This key is set post-manufacturing, so the correct functionality is never revealed to the untrusted foundry \cite{Tehranipoor2015CounterfeitAvoidance}. 

Early examples of logic locking techniques insert keyed XOR and MUX gates to corrupt the next-state logic \cite{J.A.Roy2008EPIC:Circuits,Rajendran2015FaultEncryption}. Unfortunately, these methods have been largely broken using a variety of methods, the most successful of which are SAT-based attacks  \cite{Subramanyan2015EvaluatingAlgorithms}. Researchers have attempted to increase the difficulty of mounting a SAT-based attack by inserting SAT-resilient logic blocks into the locked circuit \cite{Xie2019Anti-SAT:Locking,Yasin2016SARLock:Locking}. These techniques reduce the number of keys ruled out per attack iteration, significantly increasing the overall execution time; however, the logic blocks are susceptible to removal attacks since the circuitry is typically traceable through properties such as signal probability \cite{Yasin2017SecurityAnti-SATb}. 

Other schemes rely on circuit properties that the SAT-based attack does not model. For instance, the Cyclic Obfuscation scheme assumes SAT solvers can only handle acyclic circuits \cite{Shamsi2017CyclicCircuits}. It creates loops in the circuit's combinational logic to corrupt the SAT-based attack. Another technique, Delay Locking, adds tunable delay key gates to the design. Incorrect keys lead to setup and hold timing violations that SAT solvers cannot model by default. This comes at the cost of large delay overheads as the security scales, with an average delay overhead of 60\% at their highest security level \cite{Xie2017DelayOverproduction}. While the security of these defenses initially seems promising, when their properties are correctly formulated within SAT, they can be easily broken \cite{Chakraborty2018TimingSAT,Zhou2017CycSAT:Encryptions}. 

Yet another class of locking mechanisms is sequential logic locking. One such technique modifies the circuit's finite state machine (FSM) to require a specific input sequence to transition from the reset state to the functionally correct set of states \cite{Chakraborty2009HARPOON:Protection}. If an incorrect sequence is given, the circuit remains in a portion of the FSM with incorrect behavior. This particular scheme is susceptible to a targeted removal attack \cite{Duvalsaint2019CharacterizationATPG,Meade2017RevisitDefenses}. In general, sequential logic locking relies on limited scan access to a design. However, recent work has extended the SAT-based attack, originally developed for combinational circuits, into a model checker-based attack that assumes no scan chain access \cite{Massad2017ReverseAccess,Shamsi2019KC2:Deobfuscation}.

In this paper, we propose a novel logic locking mechanism, namely the addition of programmable phase latches within the design. The latches enable the obfuscation of functionality on multiple levels: manipulating the size and location of the circuit's state, corrupting of the circuit's timing behavior, and adding decoy logic. Our analysis shows that latch-based logic locking is SAT/model checker-based attack resistant and requires minimal overhead, enabling large amounts of insertion. The contributions of this work are the following:
\begin{itemize}
\item Introduction of latch-based logic locking, a locking technique that operates via manipulation of the phase of sequential elements 
\item A design flow for latch-based logic locking, leveraging the existing commercial synthesis tools
\item Evidence of small design overhead for a set of industrial cores and benchmark circuits from synthesis runs as well as taped out place and route runs
\item Comprehensive attack modeling of locking properties including reset, cycle delay, timing, and loop breaking constraints
\item Demonstration of resistance to current model checker-based attacks
\end{itemize}

\section{Background}
\subsection{Attacker Model}
In the characterization of the security of a locking technique, an attacker model is used to specify assumptions regarding the adversary's ability. In this paper, we assume that the adversary has access to three artifacts: the locked circuit's netlist, the timing information of the netlist, and an unlocked version of the circuit with the correct key set inside the tamper-proof memory. This unlocked circuit is referred to as the oracle. This model corresponds to a foundry that has netlist and timing information from a GDSII file and an unlocked circuit purchased on the open market. While we maintain the use of scan chains in the design, we assume that they have been disabled through use of a fuse mechanism. The implementation of fuses has been previously demonstrated and widely used in practice \cite{Shi2011Zero-maskProcesses,Wu2011InvestigationTechnology}. While additional side channel techniques may augment an attacker's capabilities \cite{wang2017probing}, we consider these to be outside the scope of the paper. This attacker model enables the mounting of a model checker-based attack described in Section \ref{mca}; however, we first describe its combinational equivalent, the SAT-based attack, in Section \ref{sa}.

\subsection{SAT-Based Attack}
\label{sa}
\begin{figure}[t]
  \centering
  \includegraphics[width=\columnwidth]{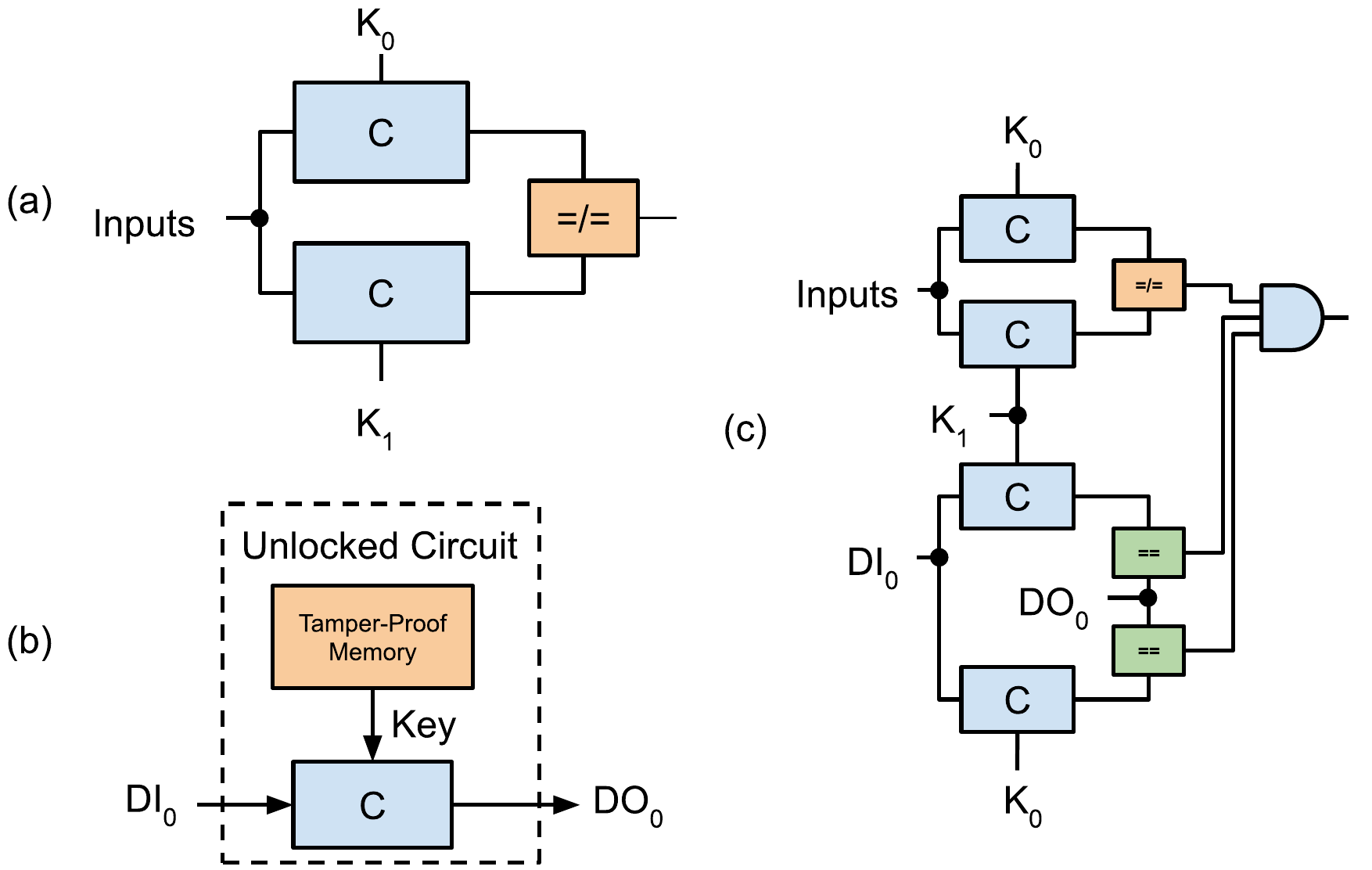}
  \caption{SAT-based attack steps: (a) Miter circuit construction consisting of two copies of the original circuit's logic (C), (b) Unlocked circuit (oracle) produces correct IO functionality (c) Addition of learned IO constraint to miter circuit}
  \label{fig:sat}
\end{figure}

 This attack uses the netlist and oracle to iteratively produce correct input-output (IO) relationships \cite{Subramanyan2015EvaluatingAlgorithms}. These relationships rule out all keys that do not produce the same behavior, narrowing the space of possible circuit functionalities. The IO relationships are efficiently learned through a three step procedure: \textbf{I.} First, a miter circuit is used to determine inputs that are guaranteed to rule out at least a single key per oracle query. A miter circuit consists of two copies of the original circuit with the inputs tied together, the key inputs kept separate, and the outputs of each connected to comparators. A diagram of the connections is shown in Fig. \ref{fig:sat}a. Additional key constraints, such as timing and loop breaking, can be conjuncted with the miter output. A SAT solver is used to find values of the shared input and keys such that the output of the miter circuit is logic 1. By construction, the solution to the SAT problem will have two different keys applied. The shared input value found by the solver is termed a differentiating input (DI). \textbf{II.} Next, as depicted in Fig. \ref{fig:sat}b, the learned DI is applied to the oracle circuit to determine the differentiating output (DO), forming an IO pair that the correct key must respect. \textbf{III.} Finally, as shown in Fig. \ref{fig:sat}c, the input and output pair is added as a constraint to the miter circuit. Now, any keys that satisfy the miter circuit will also satisfy the learned IO relationship. While each relationship is guaranteed to rule out at least one key, in practice, a larger portion of the key space is ruled out. These steps repeat until the miter circuit formulation is unsatisfiable. At this point, any key that respects all learned IO relationships will be a functionally correct key. 

\subsection{Model Checker-Based Attack}
\label{mca}
The sequential version of the SAT-based attack uses the same oracle and miter circuit, but extends the concept of a DI to a differentiating input sequence (DIS). To handle the state elements in the design, this attack substitutes the SAT solver for a model checker. The model checker is able to unroll the circuit in time, connecting the state input of one cycle to the state output of the next. Every input sequence is assumed to start from a potentially unknown, but \textit{fixed} reset state. In the same fashion as the SAT-based attack, a miter circuit is used to find DISs that produce different output sequences for different keys. These DISs are applied to the oracle and the produced outputs form IO relationships which are again encoded into the miter circuit as unrolled constraints. Aside from the reduced controllability and observability from the state elements, the most significant difference between the SAT-based and model checker-based attacks is in the termination conditions. While SAT solvers typically can exhaustively prove that no input can satisfy the miter circuit, this is a much harder task for the model checker due to the state space explosion associated with allowing the model checker to unroll the circuit indefinitely. Thus, the model checker attack relies on secondary termination conditions: a check that only one key remains and a logical equivalence check for the combinational logic fan-in into each flip-flop. For a more complete description of the model checker-based attack, we direct the reader to \cite{Massad2017ReverseAccess,Shamsi2019KC2:Deobfuscation}.


\subsection{Latch-Based Design and Retiming}
Latches are level-sensitive sequential elements that are transparent when the clock input is logic 1 and hold a sampled value when the clock is logic 0. A typical master-slave flip-flop contains two latches: a master latch with an inverted clock phase followed by a slave latch with nominal clock phase. When the clock is high, a sampled value from the last clock cycle is held by the master latch and loaded into the slave. When the clock is low, the slave latch maintains this sampled value, while the master loads a new sample. Together, the latches create an edge-triggered flip-flop, propagating the input data to the output on the rising edge of the clock. 

Through a set of transformations known as retiming, these latches can be separated into individual gates and moved through the fixed logic of a circuit \cite{Yoshikawa2004TimingLatches,Leiserson1991RetimingCircuitry}. A functionally correct retiming will maintain the cycle delay of every path through the latches. \textit{Cycle delay} is defined as the number of required clock cycles for data to propagate along a path. Thus, all paths through the original flip-flop maintain a cycle delay of 1, passing through an inverted phase latch, then a nominal phase latch. Latch retiming is used to reduce the critical path of a design, shifting the amount path delay between the positive and negative phase of the clock, as well as reduce area by merging latches. The level sensitivity of latches enables cycle sharing between the positive and negative phases of the clock cycle allowing more flexible signal arrival times. This property is often exploited in designs with tight timing constraints \cite{Wang2009IntelFPGA-synthesizable,Schelle2010IntelSynthesizable}. In this work, we use the flexible arrival time to increase the modeling difficulty of an obfuscated circuit.

\section{Latch-Based Logic Locking}
In developing a locking system, we consider maintaining circuit frequency a topmost priority. Toward this goal, we utilize latches to lock a circuit via two mechanisms: programmable path delay and programmable logic. This creates a locking system in which the critical path logic can be obfuscated while remaining largely unmodified. Both mechanisms employ decoy latch elements to cheaply create uncertainty as to the circuit's correct function. 

\subsection{Programmable Path Delay Decoys}

\begin{figure}[t]
    \centering
    \includegraphics[width=\columnwidth]{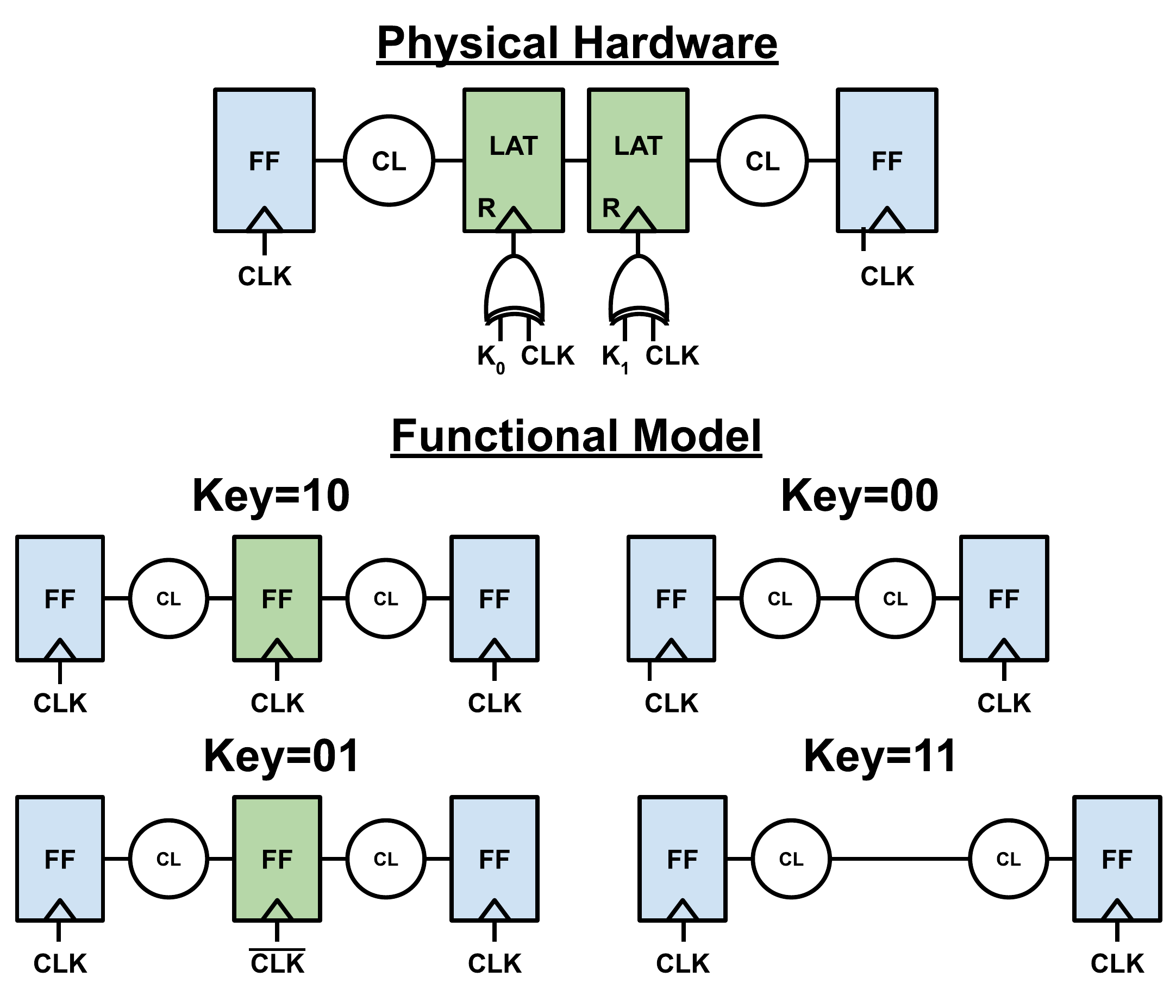}
    \caption{Example of path with programmable phase latches (LAT), flip-flops (FF) and combinational logic (CL). Both the physical hardware and functional model are shown. Depending on the key setting, the cycle and propagation delays are changed.}
    \label{fig:latch_example}
\end{figure}
As shown in Fig. \ref{fig:latch_example}, a phase-programmable latch can be created using a key to selectively feed a latch a nominal or inverted phase clock. The latches either replace existing flip-flops or are inserted as decoys on nets within the existing combinational logic. Replacing flip-flops from the original design can actually decrease the minimum clock period due to cycle sharing. If the phase of a decoy latch is picked such that the latch is clear during transitions on its input, the additional propagation delay is minimal, roughly the delay of 2 inverters. 

Depending on the key setting, these phase-programmable latches can manipulate both the cycle delay \textit{and} propagation delay along a path, thereby changing the functionality of the design via two mechanisms. \textit{Propagation delay} is defined as the amount of time necessary for a signal to travel from its launching point to capturing point. An incorrect cycle delay will change the logical behavior of the design whereas an incorrect propagation delay will cause timing violations and undetermined behavior. The pair of latches in Fig. \ref{fig:latch_example} can be set to four unique functionalities: a positive edge flip-flop (10), a negative edge flip-flop (01), a combinational path with propagation delay increased by a half-cycle (11), and the combinational path with no added propagation delay (00). As shown in Fig. \ref{fig:conceptual}, path delay decoy latches (orange) can be inserted along paths in the original combinational logic of a circuit. 

Along with their low delay impact, latches are well suited for obfuscating a circuit's path delays due to their increased complexity in modeling as compared to edge-triggered elements. Both the cycle and propagation delay added by each latch are a function of the phase of the upstream latches along a given path, thus increasing the computation required to characterize a given key setting. Additionally, because signals can be launched from the latch at any point within its clear phase, checking whether all signals are legally timed under a given key becomes significantly more challenging. We expand on the utility of both these properties in Section \ref{attack}. In summary, the reasons that make latches hard to design are the same that allow us to utilize them as a source of obfuscation.

\subsection{Programmable Logic Decoys}
\begin{figure}[t]
    \centering
    \includegraphics[width=\columnwidth,trim={1.5cm 0 1.5cm 0},clip]{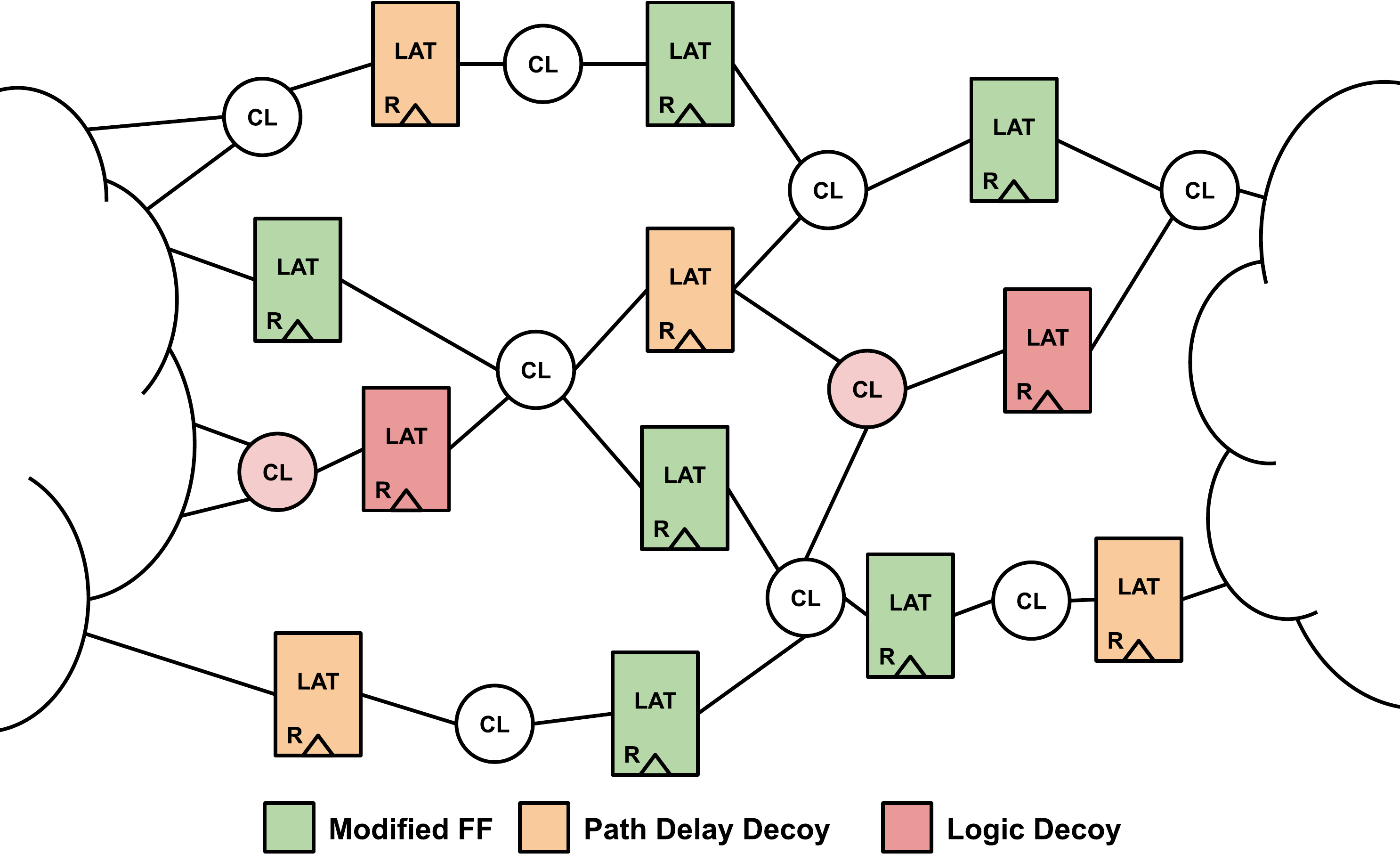}
    \caption{Conceptual view of latch-based logic locking. A set of interconnected flip-flops is converted to programmable latches with added decoy latches and logic. Each latch can operate on either clock phase, hold clear, or output constant logic 0.}
    \label{fig:conceptual}
\end{figure}

In conjunction with the manipulation of delays in the circuit, we can add decoy logic within the design to create uncertainty in the functions being computed. Keying the reset pin of the phase-programmable latches, creates a simple method of adding decoy logic. Additional latches are inserted in the design, each along with a fan-in cone of decoy logic. The latch output is combined with existing logic such that the functionality is not changed if the latch output is logic 0. When the correct key is applied to the circuit, the reset pin is forced to logic 1, thus maintaining the intended function. Example logic decoys and corresponding logic are shown in Fig. \ref{fig:conceptual}, highlighted in red. 

The added programmable logic also allows flexible insertion of latch-based logic locking. Circuits with sparse interconnection can be augmented to increase the amount of paths through the locked portion. Structures known to be harder to resolve, such as reconvergent fan-out and sequential feedback paths, can be selectively added within this additional logic. Importantly, these modifications can be added while leaving the critical path of the design unchanged.

Adding this additional key state requires an extra bit for each latch. We can make use of all four states by adding a clear mode in which the latch is held open, fixing the clock input to logic 1. This additional state can help reduce insertion overhead and concurrently increase the difficulty of modeling the system. Now the added cycle and propagation delays depend on \textit{any} upstream latch, not just the immediate fan-in latches. This locking technique forces the adversary to determine for each latch whether it is a \textbf{positive phase latch}, \textbf{negative phase latch}, \textbf{delay decoy}, or \textbf{logic decoy}.

\subsection{Insertion Flow}
\begin{figure}[t]
    \centering
    \includegraphics[width=\columnwidth]{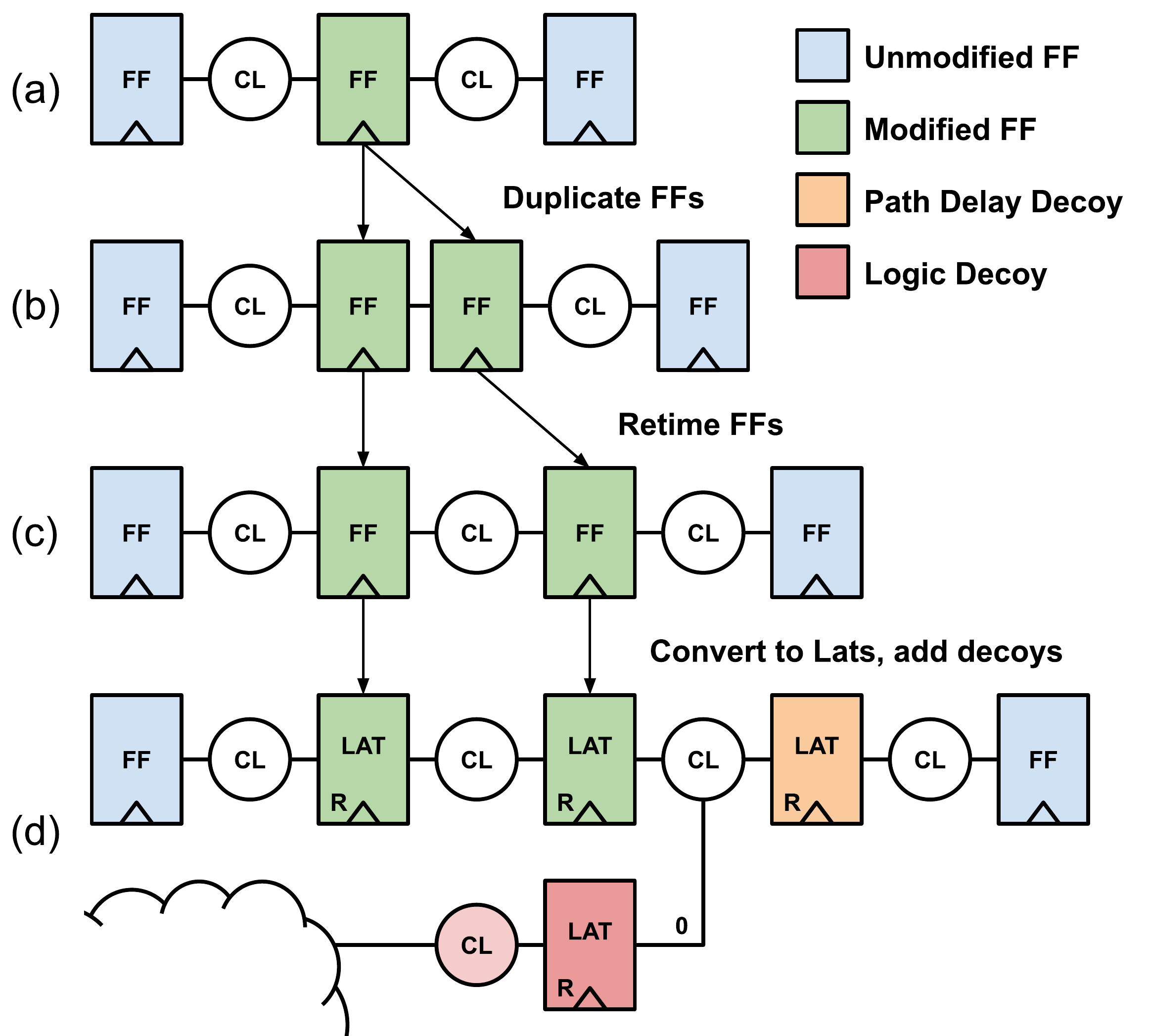}
    \caption{Flip-flop to latch conversion and decoy addition. A flip-flop is duplicated, retimed, and converted to latches. Two types of decoy latches can be added to the paths in the fan-in and fan-out cones of the latches. This example shows a single flip-flop, however in practice an interconnected group is converted.}
    \label{fig:insertion}
\end{figure}

\textbf{I.} The first step of the insertion flow is the selection of a set of interconnected flip-flops within the original netlist. When converted to phase-programmable latches, the ideal set of flip-flops will have many possible functional states that are also hard to rule out. Many possible functional states corresponds to a set of flip-flops with many paths through the group; more interconnections will create more unique paths. Additionally, we would like to minimize the controllability and observability into the group, thereby making each scenario harder to rule out. Typically, this corresponds to fewer paths in and out of the group. This task can be naturally modeled as community finding in graphs. We adopt a flow-based community finding algorithm to produce appropriately sized groups \cite{Pares2017FluidAlgorithm}. From a sample of these groups, the group with the lowest cumulative delay is selected, allowing low overhead for decoy insertion. 

\textbf{II.} The selected group of flip-flops is subsequently converted into latches and retimed. Standard synthesis tools do not support latch retiming, however their ability to retime flip-flops can be leveraged with a previously established procedure \cite{Yoshikawa2004TimingLatches,Chinnery2004AutomaticASICs}. First, each latch of the original flip-flop is replaced by a flip-flop, depicted in Fig. \ref{fig:insertion}a-b. Then, the flip-flops representing the positive phase latches are held fixed, while the flip-flops representing negative phase latches are retimed. This is then repeated, fixing the negative phase latches and retiming the positive, shown in Fig. \ref{fig:insertion}b-c. These flip-flops are then replaced by their respective latch counterparts, depicted in Fig. \ref{fig:insertion}c-d. 

\textbf{III.} Next, as shown in Fig. \ref{fig:insertion}d, the two types of decoy latches are inserted into the collective fan-out and fan-in cone of the group. Path delay decoys are randomly inserted on nodes with slack greater than the delay through a clear latch. To add logic decoys, cones of random logic driven by subsets of the latches are created. These cones are connected to a latch which fans out to another subset of the latches. The fan-out logic is created such that if the latch is controlled to constant logic 0 through the reset pin, there will be no effect on the downstream original logic. 

\begin{figure}[t]
    \centering
    \begin{minipage}[c]{0.4\columnwidth}
        \centering
        \includegraphics[width=.5\columnwidth]{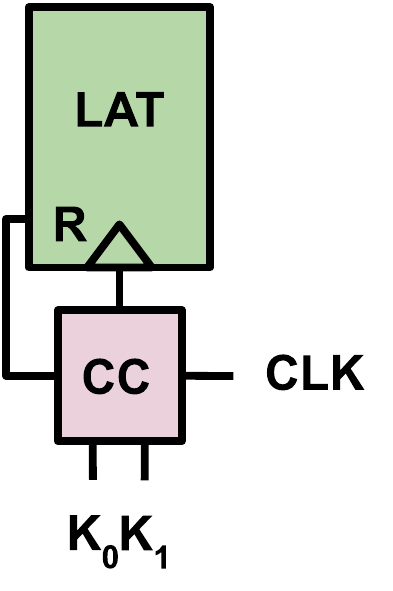}
    \end{minipage}
    \begin{minipage}[c]{0.5\columnwidth}
        \centering
        \begin{tabular}{ cc|cc }
         \hline
          \textbf{Key0} &  \textbf{Key1} &  \textbf{R} &  \textbf{CLK}\\
         \hline
         \rowcolor[HTML]{EA9999}
         0 & 0 & 1 & 0\\
         \rowcolor[HTML]{B6D7A8}
         0 & 1 & 0 & CLK\\
         \rowcolor[HTML]{B6D7A8}
         1 & 0 & 0 & !CLK\\
         \rowcolor[HTML]{F9CB9C}
         1 & 1 & 0 & 1\\
        \end{tabular}
    \end{minipage}
    \captionlistentry[table]{A table beside a figure}
    \label{tbl:latch_clock}
    \captionsetup{labelformat=andtable}
    \caption{Diagram and truth table of latch clock and reset control signals. Control circuitry (CC) which determines the latch function is connected to the clock and reset pins. The truth table's row colors correspond to the associated latch type.}
    \label{fig:latch_clock}
\end{figure}
\textbf{IV.} Finally, as depicted in Fig. \ref{fig:latch_clock}, the logic to control the clocking and reset of all the latches is added. Each latch is controlled with two key bits. The four resulting states correspond to a clear buffer, a positive phase latch, a negative phase latch, and a constant logic 0. The truth table for the logic is shown in Table \ref{tbl:latch_clock}. During timing optimization and analysis, this logic is constrained to the correct key setting so the tool can target proper function. 
To verify the locking process did not corrupt the functionality, the correct key is fixed and a sequential equivalence check between the modified and original design is run. Additionally, timing-annotated functional simulations can provide further assurance that the timing constraints have been properly specified.

\subsection{Design for Testability}
A primary concern when developing an IC is testability. Design flows typically employ scan-based flops to serially load test vectors into the circuit; however, this design for testability (DFT) mechanism is not directly compatible with a latch-based design. In order to maintain test coverage of our locked circuits, we utilize scannable test points inserted at every latch in the design. Through the use of two muxes (only one on the functional path) and a scan flip-flop, the output of each latch can be observed and controlled to an arbitrary value, emulating a full scan methodology. Faults in the added clock tree logic can be covered with additional, shared observation points.

We implement this initial version of the test infrastructure to demonstrate that the latch-based logic locking does not significantly impact testability; the results are shown later in Section \ref{test_section}. This DFT strategy as outlined maintains low delay impact, but sacrifices area for ease of implementation. This strategy can be improved using known techniques explicitly designed for latches, namely level-sensitive scan design (LSSD). Using only a mux and an additional latch, the same test coverage can be achieved; however, this implementation is left for future work. 

\section{Attacking Latch-Based Logic Locking} \label{attack}
\subsection{Key Constraints}
Given the targeted attacker model, the adversary has access to a locked netlist with timing information. Assuming a typical design flow, the adversary can constrain the key space to rule out scenarios wherein combinational loops or timing violations are created. The process of forming each constraint type is explained below. 

In the latch-based logic locking insertion flow, potential combinational loops are likely created either when converting an original flip-flop to keyed latches or when inserting programmable logic decoys. A potential combinational loop exists if there is a cyclic path through the latches. The savvy adversary can form key constraints to ensure that along each of these cyclic paths there is at least one phase transition. These loop breaking constraints can be found by tracing paths from each latch back to itself, encoding that not all encountered latches are on the same phase or at least one of the latches is held in reset. 

Similarly, the adversary can utilize netlist timing information to rule out additional key combinations. 
Assuming every path in the correct circuit meets timing and that the clock period is known within some bound, key constraints can be created to exclude violating configurations. 
The clock period can be estimated via analysis of unkeyed paths or directly obtained through analysis of an oracle circuit. 

As previously mentioned, the potential for cycle sharing makes timing latches significantly more flexible than flip-flops. 
Ruling out \textit{guaranteed} incorrect keys can be done using the maximum amount of cycle sharing which limits the propagation delay through any contiguous set of latches on the same phase to at most the clock period, $t_{period}$. 
Additionally, the propagation delay through any contiguous set of latches with only 1 transition in phase is limited to $1.5\times t_{period}$. These constraints are generated considering every path containing a latch. 
Along each such path, every delay window of length $t_{period}$ and $1.5\times t_{period}$ is constrained to have at least 1 and 2 transitions respectively. 

\subsection{Key Equivalence Conditions}
\begin{figure}[t]
    \centering
    \includegraphics[width=.9\columnwidth]{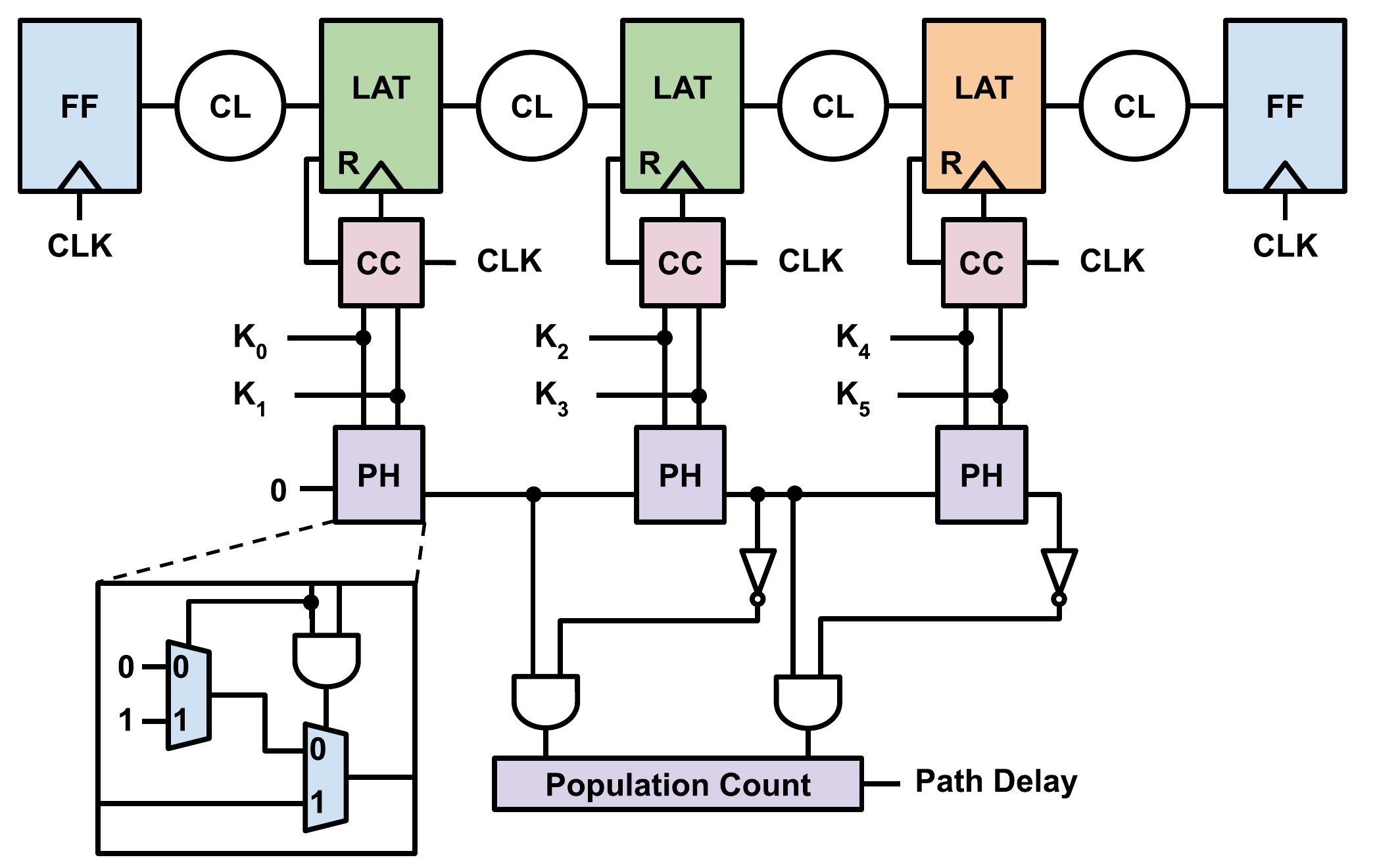}
    \caption{Example path cycle delay circuit, helping create equivalence classes for keys. The circuitry counts the number of cycle delays along a path through the keyed latches. CC represents clock and reset control circuitry from Table \ref{tbl:latch_clock}. PH represents circuitry that determines the phase of the corresponding latch. Population count is used to determine the number of transitions.}
    \label{fig:equiv}
\end{figure}
As previously mentioned, critical to the success of model checker-based attacks is grouping functionally equivalent keys. This entails deriving a function that maps a given key to a point in the circuit's functionality space. Once this function is established, if a particular key's functionality can be ruled out, all other keys mapping to the same functionality can be ruled out as well. This function is used in the termination condition of the model checker-based attack. In latch-based logic locking, there are two forms of programming functionality: programmable cycle delay and programmable logic; both must be considered to properly determine key equivalence. (The propagation delay is also modified, but the potential timing violations this causes can be considered separately.)

Programmable cycle delay is closely related to the process of retiming which moves registers through combinational logic while maintaining functionality. If two circuits with the same combinational logic are retimings of each other, than they are equivalent. Equivalence conditions can be established using a result borrowed from circuit retiming. The authors in  \cite{Leiserson1991RetimingCircuitry,Leiseron1983OptimizingSystems.} establish that any retimed circuit will have equal cycle delays along any path through the logic. Likewise, equivalent functionalities of a design with programmable cycle delays will maintain this same relationship.

This relationship can be encoded into a function by creating a cycle delay counting circuit as shown in Fig. \ref{fig:equiv}. The circuit consists of logic to determine the phase of each latch within the design (PH) and logic to count the number of phase transitions. The PH logic outputs either positive phase, negative phase, or feeds though the upstream latch's phase depending on the key. The number of latch phase transitions is calculated using a population count circuit. A cycle delay counting circuit is built for all paths originating in the unmodified portion of the design tracing through the latches and for any path from a latch back to itself.

Additionally, latch-based logic locking can change logic functionality by holding a latch's reset pin, creating a constant logic 0 on the output. Different logic functionalities can be detected with a combinational equivalence check of the logic fanning into each latch that is not held clear or at reset. Whereas the previous equivalence condition checks that the data flows through the logic at equivalent rates, this equivalence condition determines that the functionality of the logic is the same. 

These two equivalence conditions are merged to form a function that maps each key to a specific functionality. The resulting circuits from two different keys can be determined to be equivalent in the following manner. Every path must have the same cycle delay count or be broken by a reset latch in both circuits. Additionally, the input function to every output, flip-flip, or latch not held at reset or clear must be the same.

\begin{figure}[t]
  \centering
  \includegraphics[width=.9\columnwidth]{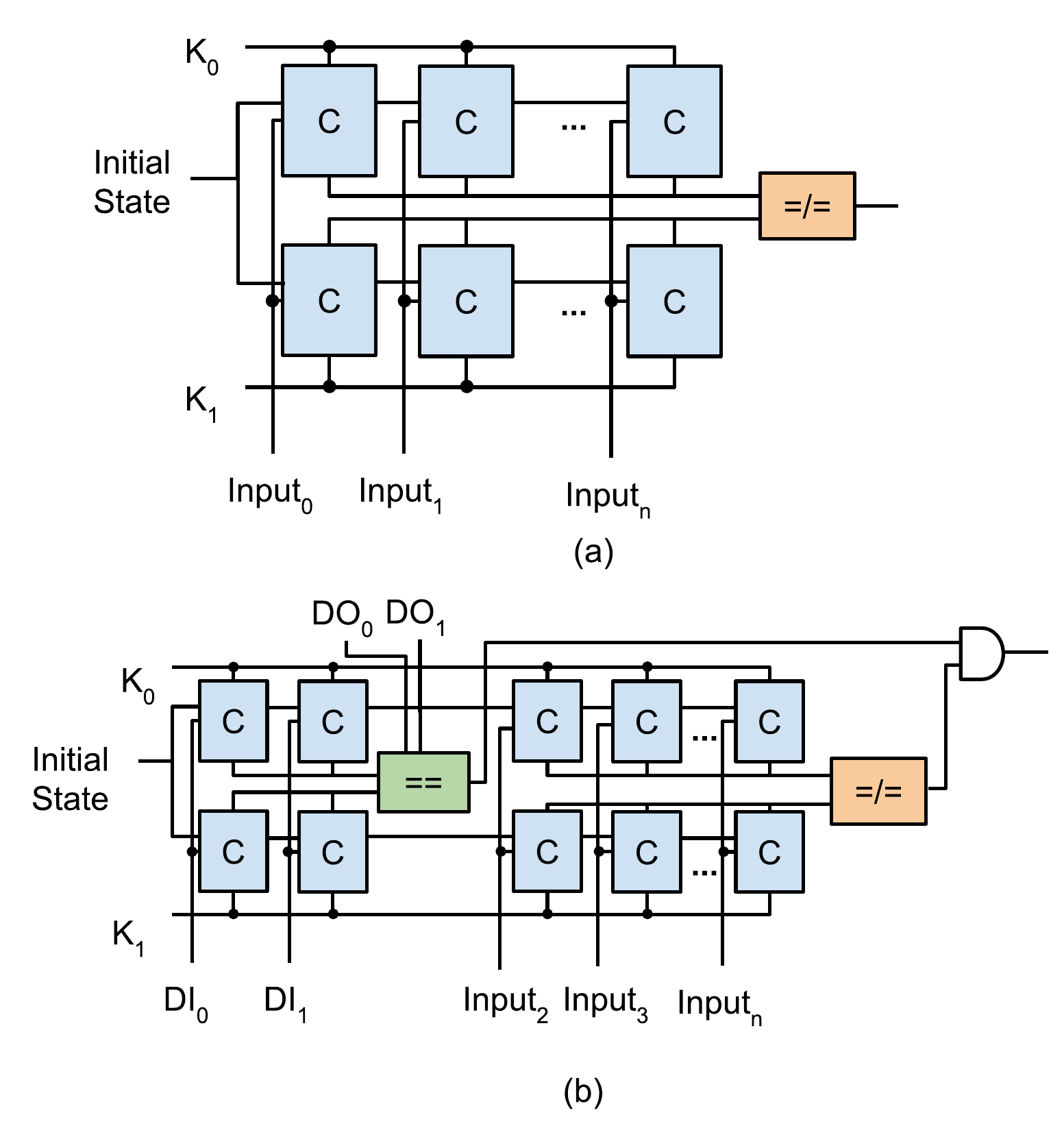}
  \caption{(a) Miter circuit used in our modified model checker-based attack. The free variables of the system are the two key inputs, the initial state of all latches, and the inputs at every unrolled cycle. The circuit is unrolled by connecting the input of each state element from a cycle to the output of the same state element of the next cycle. (b) Addition of a learned IO constraint to the miter circuit. In this case, the inputs of the first two cycles are no longer free variables.}
  \label{fig:unroll}
\end{figure}

\subsection{Incompatible Reset Assumptions of Previous Attacks}
If the attacker has access to an oracle circuit, the previously established key constraints and equivalence conditions can be integrated into a model checker-based attack. In describing such an attack we must first address an incompatibility with model checker-based attacks from previous works \cite{Massad2017ReverseAccess,Shamsi2019KC2:Deobfuscation}. The issue stems from the assumption that a circuit's state is able to be reset to a potentially unknown, but \textit{fixed} value. In many circuits this assumption is untrue. As a matter of fact, IC designers frequently include flip-flops with no reset in their designs to save both area and power. Often only a core portion of sequential elements are set to a fixed state during reset, leaving other sequential elements in unknown states. The sequential elements that are uninitialized will eventually be set to a known state during the circuit's operation. Examples of this behavior include the register file of many processors or the datapath registers of AES. 

Previous attacks use this fixed reset assumption to repeatedly find differentiating input sequences starting from the reset state. If this assumption is violated, the learned IO relationships will be corrupted because of their dependence on unfixed, unmodeled variables. Since many circuits violate this assumption and latch-based logic locking adds state that potentially has an unfixed reset, we generalize the attack to account for this behavior. 
\begin{table}[t]
    \centering
    \begin{tabular}{ cccc }
     \hline
     \textbf{Circuit} &  \textbf{FFs} &  \textbf{Gates} &  \textbf{Frequency(GHz)}\\
     \hline
     IIR & 650 & 6308 & 0.79\\
     FIR & 568 & 6854 & 0.88\\
     DES3 & 135 & 3023 & 1.42\\
     AES & 530 & 12088 & 1.79\\
     OR1200 & 3942 & 20252 & 0.77\\
     s9234 & 145 & 902 & 2.27\\
     s13207 & 597 & 2368 & 1.78\\
     s15850 & 513 & 2691 & 1.56\\
     \hline
    \end{tabular}
    \caption{Characteristics of each considered benchmark circuit}
    \label{tbl:designspecs}
\end{table}
\subsection{Model Checker-Based Attack with Non-Deterministic Reset}
The generalized model checker-based attack utilizes an iteratively unrolled miter circuit as depicted in Fig. \ref{fig:unroll}a. The circuit’s free variables are the separate key inputs, shared inputs, and shared the initial state. The key constraints and equivalence functions are instantiated along with the miter circuit such that the circuits produced by $K_0$ and $K_1$ do not have paths that explicitly violate timing, do not contain combinational loops, and have different functionalities. The first cycle input is constrained such that reset is active; in subsequent cycles, the tool is free to control reset. The model checker is given the task of finding an input sequence that satisfies the miter circuit, thereby producing a DIS. The tool searches for the sequence, unrolling the circuit as needed. Since the latches operate on both phases of the clock, every clock cycle is modeled with two copies of the circuit.

Once a DIS is found, the oracle is used to find the correct output. The input sequence is run on the oracle, capturing the output and holding the clock steady on the last cycle. The produced IO relationship is encoded as a constraint on the miter circuit as depicted in Fig. \ref{fig:unroll}b. Now, satisfying the miter circuit entails finding two keys and an initial state that agree with the learned IO sequence thus far as well as extending the input sequence to distinguish additional keys. Since the state of the system is maintained between DISs, one contiguous IO relationship is formed, avoiding the issues with previous attacks.

After every newly encoded IO constraint, a termination check is run. Using a model checker run bounded to the length of the IO sequence thus far, the tool attempts to find two keys and initial state that agree with the IO sequence, meet the key constraints, and are functionally different. If the tool cannot find two such keys, the attack terminates. Otherwise this process is repeated, extending the length of unrolling. 

This attack creates a circuit that generates functionally correct key, initial state pairs. However, the initial state of the system cannot be directly set, since it is determined by the random settling of the state elements. The generated key is only guaranteed to have correct IO behavior when paired with its specific initial state, thus when applied to the circuit it may produce spurious behavior. To avoid this, an attacker attempting to overproduce the design, can modify the reset sequence to explicitly set the initial state.

\section{Experimental Results}
\begin{figure}[t]
  \centering
  \includegraphics[width=\linewidth]{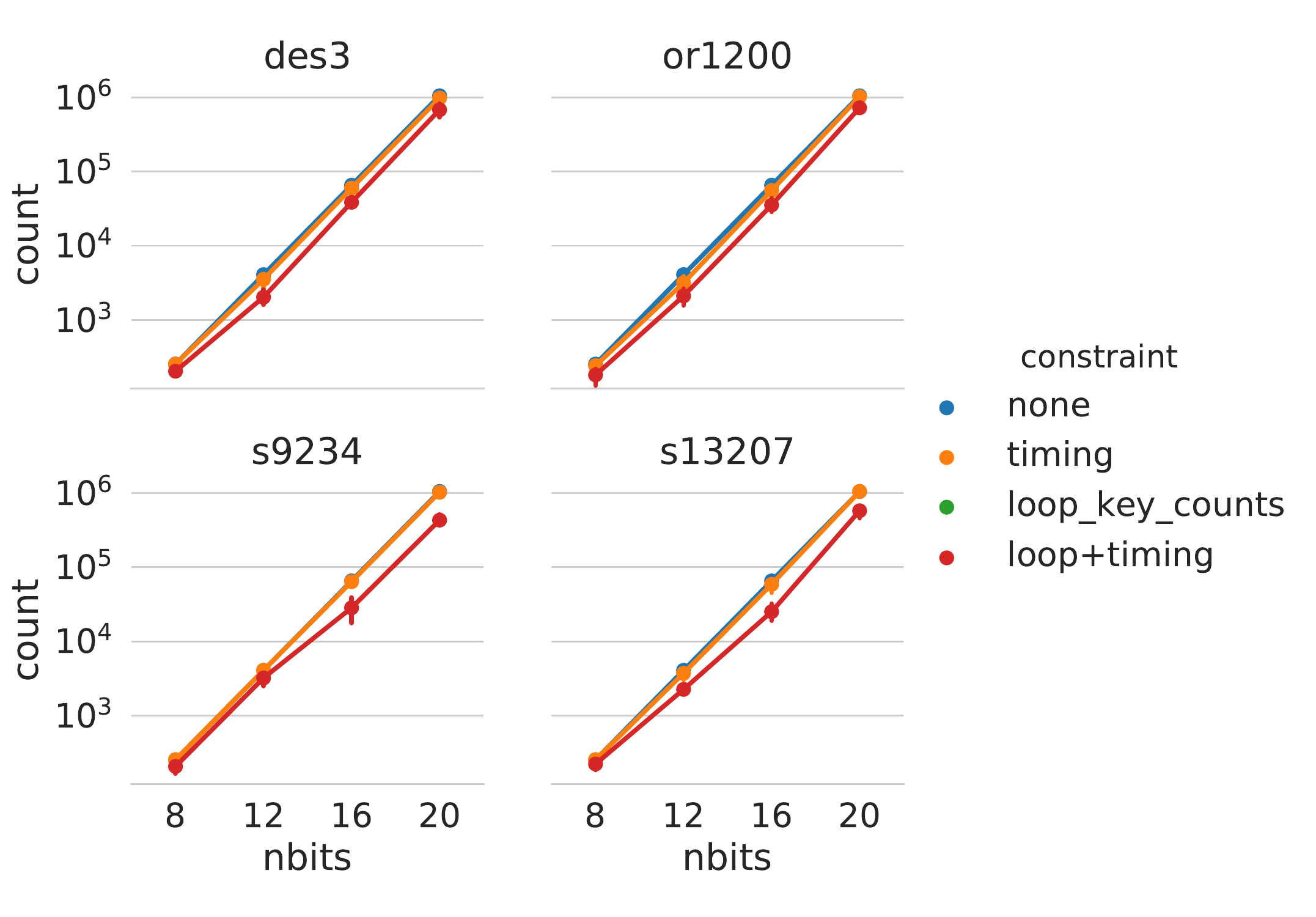}
  \caption{Enumeration of keys that satisfy timing and combinational loop breaking constraints}
  \label{fig:constraint}
\end{figure}
We demonstrate the attack resistance and design overheads associated with latch-based logic locking on two sets of open source circuits: the Common Evaluation Platform and ISCAS89 benchmarks. All circuits were synthesized using a modern commercial standard cell library in a 22nm process. The characteristics of each design are displayed in Table \ref{tbl:designspecs}. As the latch-based logic locking insertion has an element of randomness depending on the inserted decoy logic, all experiments are run with 5 samples per circuit, bit-count pair. We target a 50\% decoy latch to real latch ratio, however the actual value depends on the circuit due to the variability of the retiming process.  

For modifying the netlists and synthesis, we used Cadence Genus. For place and route runs, we used Cadence Innovus. All synthesis and place and route runs target the maximum obtainable frequency. For assessing ATPG overheads, we use Mentor Fastscan. Finally, for the model checker based attack, we use Cadence JasperGold. All attacks are run with parallelization enabled on a server with 750 GB RAM and 16 2.1GHz cores. JasperGold dynamically optimizes the execution strategy (e.g., number of parallel threads, solver type, and memory allocation) depending on the problem. 

\subsection{Attack Results}
We first assess the impact of constraining the key space with the combinational loop breaking and timing constraints. We generate the constraint functions, sweeping the number of key bits from 8 to 20. Then, we count the number of keys that satisfy each constraint function. In Fig. \ref{fig:constraint}, we plot the counts of both types of constraints as well as the counts of keys in the intersection of the constraints. The resulting trend clearly demonstrates that the portion of the keys ruled out by the constraints is minimal, with the valid key count remaining exponential in the number of bits.

Next, we run the model checker-based attack on each circuit, sweeping the number of key bits from 8 to 32. Past 32 bits of obfuscation, we begin to see failures in the equivalence function generation as a result of the exponential number of paths that must be constrained. The attack is run with an initial timeout of 4 hours. The average execution time, $\bar{t}_{exec}$ (s), and number of timeouts, $N_{tmo}$, are reported in Table \ref{tbl:attack}. To validate the key returned by the attack, we fix one miter key input to the correct value while setting the other key input and initial state to the values returned by the attack. An unbounded model check is run to verify the miter cannot be satisfied. Fixing each key significantly reduces the state space and the tool is able to determine sequential equivalence.

\begin{figure}[t]
    \centering
    \includegraphics[width=\columnwidth]{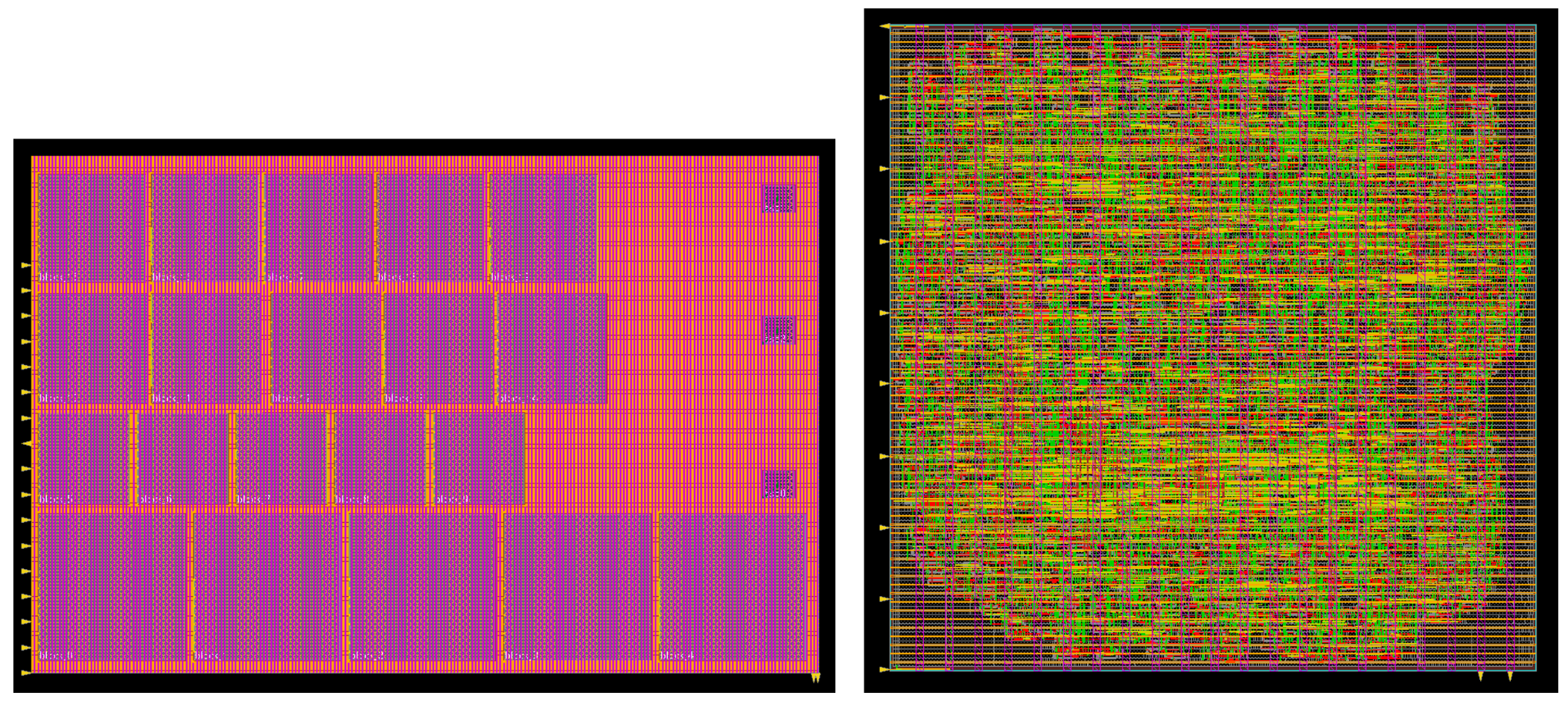}
    \caption{GDSII view of the taped out circuits. (Left) From top to bottom: FIR, IIR, DES3, AES; from left to right: original, 32 bits, 64 bits, 128 bits, 256 bits. (Right) Block view of IIR locked with 256 bits}
    \label{fig:tapeout}
\end{figure}
\begin{table*}[t]
    \centering
    \begin{tabular}{ccc cc cc cc cc cc cc}
     \hline
     & \multicolumn{2}{c}{8 bits} & \multicolumn{2}{c}{12 bits} & \multicolumn{2}{c}{16 bits}& \multicolumn{2}{c}{20 bits}& \multicolumn{2}{c}{24 bits}& \multicolumn{2}{c}{28 bits} & \multicolumn{2}{c}{32 bits} \\
    Circuit & $\bar{t}_{exec}$ & $N_{tmo}$ & $\bar{t}_{exec}$ & $N_{tmo}$& $\bar{t}_{exec}$ & $N_{tmo}$& $\bar{t}_{exec}$ & $N_{tmo}$& $\bar{t}_{exec}$ & $N_{tmo}$& $\bar{t}_{exec}$ & $N_{tmo}$& $\bar{t}_{exec}$ & $N_{tmo}$\\ 
     \hline
     DES3 & 2919 & 1 & 2915 & 1 & 3039 & 1 & 8809 & 3 & 8902 & 3 & 8857 & 3 & - & 5\\
    FIR & 176 & 0 & 157 & 0 & 5915 & 2 & 3078 & 1 & 5932 & 2 & 9061 & 3 & 11590 & 4\\
    IIR & 789 & 0 & 349 & 0 & 3245 & 1 & 4820 & 1 & 3268 & 1 & 6172 & 2 & 11627 & 4\\
    AES & 2308 & 0 & 270 & 0 & 6051 & 2 & 6223 & 2 & 11764 & 4 & 8941 & 3 & - & 5\\
    OR1200 & 12603 & 4 & 12839 & 4 & - & 5 & - & 5 & - & 5 & - & 5 & - & 5\\
    s9234 & - & 5 & 12115 & 4 & - & 5 &  11834 & 4 & 11742 & 4 & 11841 & 4 & - & 5\\
    s13207 & 11713 & 4 & 9055 & 3 & 11883 & 4 & 12545 & 4 & - & 5 & - & 5 & - & 5\\
    s15850 & 6316 & 2 & - & 5 & - & 5 & - & 5 & - & 5 & - & 5 & - & 5\\
     \hline
    \end{tabular}
    \caption{Model checker-based attack average execution times, $\bar{t}_{exec}$ (s), and number of timed out runs, $N_{tmo}$, from 5 iterations.}
    \label{tbl:attack}
\end{table*}
\begin{figure*}[t]
  \centering
  \begin{minipage}[c]{0.9\textwidth}
        \centering
  \includegraphics[trim={0 0 5.5cm 0},clip,width=\textwidth]{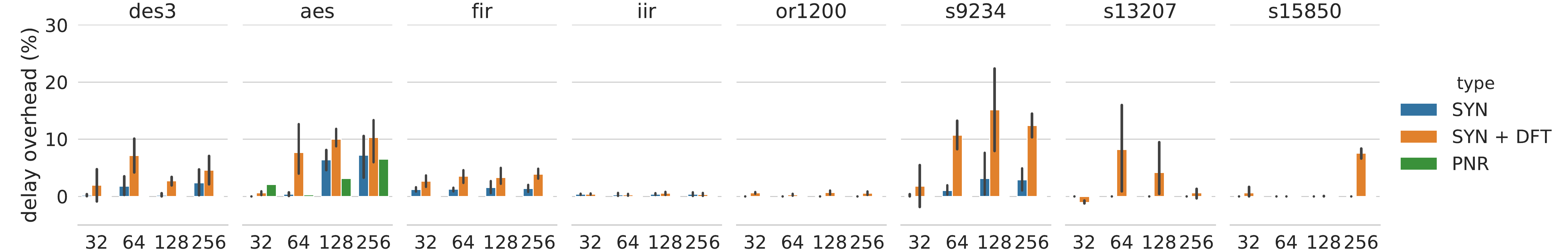}\\
   \vspace{3.00mm}
  \includegraphics[trim={0 0 5.5cm .7cm},clip,width=\textwidth]{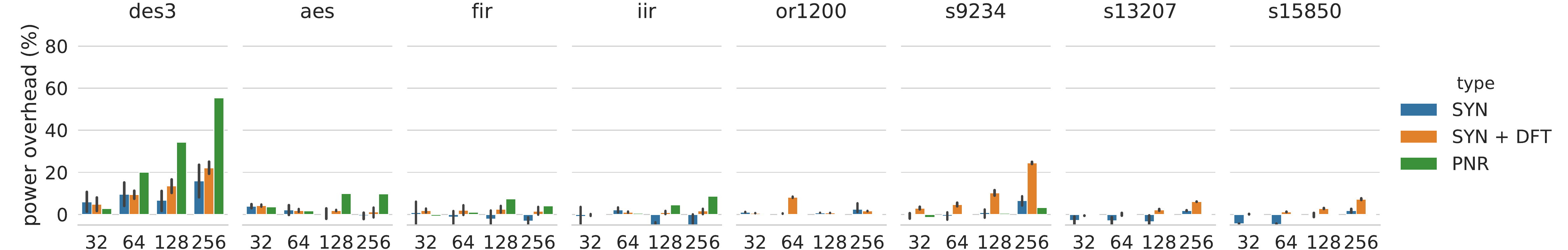}\\
   \vspace{3.00mm}
  \includegraphics[trim={0 0 5.5cm .7cm},clip,width=\textwidth]{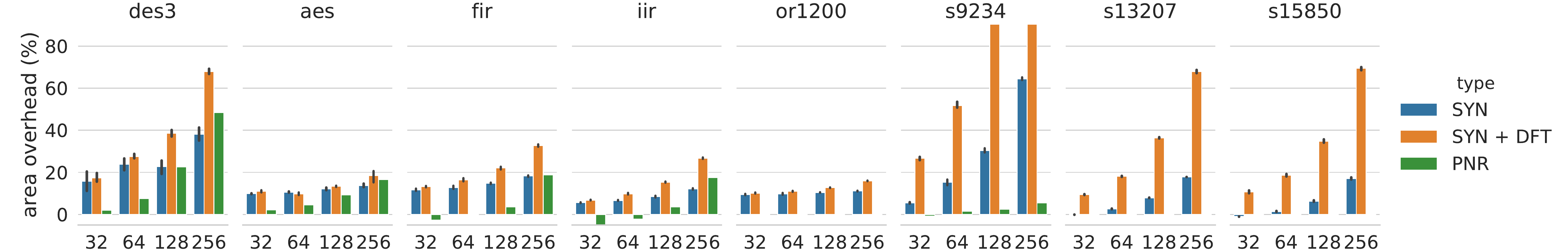}
  \end{minipage}
  \begin{minipage}[c]{0.09\textwidth}
        \centering
  \includegraphics[trim={42cm 0 0 0},clip,width=\textwidth]{images/delay_overhead-eps-converted-to.pdf}
  \end{minipage}
  \caption{Delay, power, and area overheads of latch-based logic locking normalized to the original design. The results are shown with and without scan chain logic colored orange and blue respectively. Place and route results (n=1) are shown for DES3, FIR, IIR, and AES in green.}
  \label{fig:overhead}
\end{figure*}
Attack runs that timeout have remaining non-equivalent keys, validated by the sequential equivalence check. At 32 bits, only 2 instances out of 40, one FIR and one IIR circuit, do not timeout. The exact scaling is not evident from the data as the timeout clips the true average values. However, we have run many attacks past the 24 hour mark and do not observe convergence in any additional circuits. The attack seems limited by the state space explosion. For example, we see that in 24 hours the model checker is only able to fully explore just 13 cycles of the AES circuit. It should be noted that this is just the lower bound as JasperGold utilizes several parallel execution engines attempting to find traces both at the current cycle bound, but also well beyond it. The lack of termination suggests that differences in behavior that are buried deep in the FSM are unlikely to be discovered. Thus, these results show resistance to this attack with a low number of key bits. The results also suggest circuit regularity may impact attack performance. OR1200, s9234, s13207, and s15850 are less regular compared to FIR, IIR, AES, and DES3 and show greater attack strength with most instances of these circuits timing out even with as few as 12 bits.

Under the returned keys, timing violations are likely to remain from incorrect propagation delays. The keys that are ruled out via the timing constraints are just the keys that will violate timing under any possible cycle sharing. After running the model checker-based attack, the adversary must still find a key that passes timing. This can be done though recombining the latches into flip-flops and retiming. Additionally, the key must be combined with the final state of the attack to guarantee matching functionality. Keys applied without changing the initial state may still produce incorrect behavior; assigning the respective initial state will force the circuit to the part of the FSM that matches the oracle. 

\subsection{Power, Performance, Area Overhead}

We obtain power, performance, and area (PPA) overheads as follows. First, the maximum frequency of the design is found via a synthesis run targeting an unattainably high frequency. The synthesis tool produces a critical path that determines the maximum frequency, but with excessive buffering. The synthesis process is then \textit{rerun from scratch targeting this maximum frequency}. This result serves as the baseline implementation to which various amounts of locking are applied and to which the results are normalized. Each circuit's baseline is keyed with 32, 64, 128, and 256 bits of latch-based logic locking, targeting the same maximum frequency. The resulting overheads are designated by the blue bars in Fig. \ref{fig:overhead}. Additionally, we assess the overhead after scan chains and test points have been inserted; these results are normalized to the baseline design with full scan, and shown in orange. Finally, for a subset of the circuits (FIR, IIR, DES3, AES), we have taken the designs through place and route, submitting them for fabrication. The top level diagram of the taped out IC and a block view of one circuit are shown in Fig. \ref{fig:tapeout}. We report results from taped out place and route runs in green. 

Without DFT, we see virtually no impact on delay or power. At 256 bits the average overheads are 1.8\%, 2.5\%, and 24.1\% for delay, power, and area respectively. Adding DFT, results in a slight increase in the delay and power overheads along with a significant jump in area cost. Now the average overheads at 256 bits are 5.0\%, 8.3\%, and 61.7\% respectively. As previously noted, this area overhead will be significantly reduced when using LSSD techniques as opposed to the current over abundance of area-expensive test points. Reducing the area overhead will also likely reduce the power and delay overheads as the length of routes should decrease. The place and route overheads remain largely consistent with the synthesis data. One notable difference is an increase in power of DES3. This is a result of the clock tree's power being included after place and route. While not a significant change for other circuits, since DES3 has a smaller proportion of flip-flops, the impact of many nodes switching with an activity factor of 1 is greater. 

In general there exists a good balance between the benefits of cycle sharing from replacing original flip-flops with latches, and the added delays from inserting decoys. These results show that the amount of locking can be greatly increased before significant overhead costs are observed.

\subsection{Testing Overhead}
\label{test_section}
\begin{figure}[t]
    \centering
    \includegraphics[width=\columnwidth]{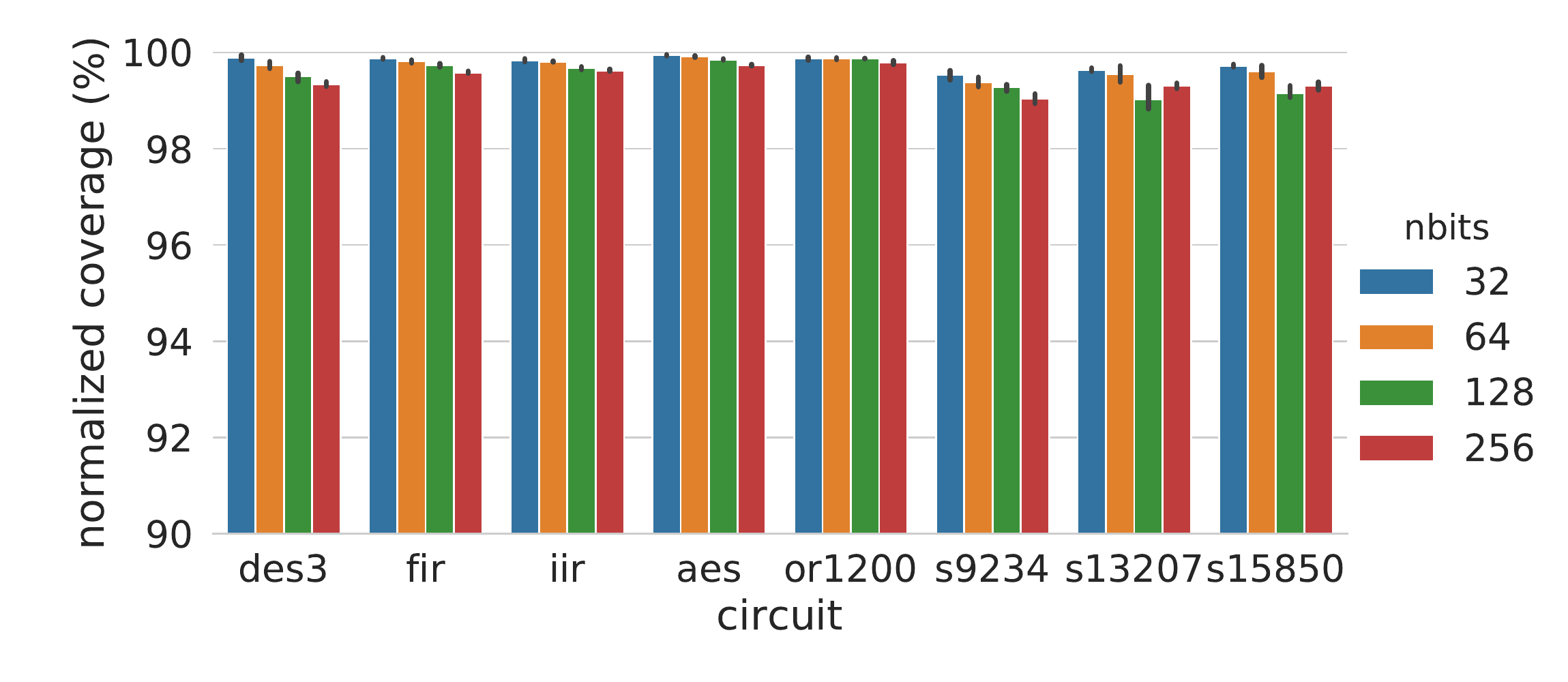}
    \caption{Normalized SSL test coverage for circuits locked with latch-based logic locking}
    \label{fig:coverage}
\end{figure}

Next, we analyze the impact of latch-based logic locking on single stuck line (SSL) fault coverage. For all circuits, we run ATPG to generate tests for all SSL faults. The resulting coverages, normalized to the original circuit with full scan, are displayed in Fig. \ref{fig:coverage}. We see an average coverage reduction of 0.2\%, 0.3\%, 0.5\% and 0.5\% for 32, 64, 128, and 256 bits, respectively. Across all designs and iterations we see a worst case reduction of 1.2\%. Assuming that design security is a top priority, these coverage overheads are likely acceptable. Furthermore, inspecting the remaining faults reveals a significant portion of the remaining faults are located along the modified clock path to the latches. Thus higher coverages are likely obtainable using the same DFT techniques used in clock gating. 
\section{Discussion}

As we have demonstrated, latch-based logic locking is resistant to current model checker-based attacks and the low overhead allows the amount of obfuscation to be increased significantly without performance loss. A key aspect of this security results from the state space explosion in which every cycle of unrolling increases the number of possible vectors by factor of $2^{inputs}$. Any successful attack must handle this problem and potential improvements to latch-based logic locking may exacerbate it. 

As seen from previous locking techniques, improved attack modeling can significantly reduce the difficulty of deobfuscation. While we have modeled our technique with significant detail, better equivalence models and constraints may exist. 
One solution to this threat is to increase the difficulty in modeling through the violation of typical design constraints. In this paper we consider two key constraints, timing and combinational loop breaking. Explicitly breaking these assumptions by adding timing violations or combinational loops in the correct design will reduce the number of keys that can be ruled out and make forming each constraint significantly more complex. Another method of increasing modeling complexity is combining latch-based logic locking with other logic locking schemes, forcing the attacker to understand and model multiple interacting circuit transformations. 

Additionally, the best structure and insertion technique for decoy logic has yet to be determined. Our current method, randomly forming logic cones from existing signals, does not add significant signal skew. Ideally this property would be kept, maintaining low traceability, while also increasing the hardness of SAT deobfuscation. We leave this and the preceding directions for future work.

\section{Conclusion}
In this work, we have introduced Latch-Based Logic Locking, a novel technique that obstructs IP theft. By manipulating the flow of data and inserting additional logic without effecting the critical paths of a design, this scheme is able to produce designs with both minimal overhead and strong attack resistance. Furthermore, this technique is compatible with industry standard tools. We have demonstrated these properties by constructing a locking flow and taking it through tape out, presenting the overhead on realistic circuits, and executing a state-of-the-art attack algorithm.

\section*{Acknowledgments}
This work was supported in part by the Defense Advanced Research Projects Agency under contract FA8750-17-1-0059 “Obfuscated
Manufacturing for GPS (OMG)” and Honeywell Federal Manufacturing \& Technologies, LLC under contract A023646.

\bibliographystyle{unsrt}  
\bibliography{mendeley,additional} 
\end{document}